\documentclass{ws-ijmpd}
\usepackage{amsfonts}
\DeclareSymbolFont{lettersA}{U}{txmia}{m}{it}
\DeclareMathSymbol{\muup}{\mathord}{lettersA}{22}

\begin{document}

\markboth{G. BAO et al.} {Improved Simulation of the Mass Charging
for ASTROD I}

%
\catchline{}{}{}{}{}
%

\title{Improved Simulation of the Mass Charging
for ASTROD I
}

\author{GANG BAO$^{1,2,3}$, WEI-TOU NI$^{1,3}$, D. N. A. SHAUL$^{4}$, H. M. ARAUJO$^{4}$, LEI LIU$^{1,2}$, T. J. SUMNER$^{4}$}
\address{1. Center for Gravitation and Cosmology, Purple Mountain Observatory, Chinese Academy of Sciences,
Beijing West Road No.2,
Nanjing, 210008, China\\
2. Graduate University of Chinese Academy of Sciences, Beijing, 100049, China\\
bgastro@pmo.ac.cn\\
3. National Astronomical Observatories, Chinese Academy of Sciences,
Beijing, 100049, China\\
4. Department of Physics, Imperial College London, London, SW7 2BZ,
UK }

\maketitle

\begin{history}
\received{Day Month Year} \revised{Day Month Year} \comby{}
\end{history}

\begin{abstract}
The electrostatic charging of the test mass in ASTROD I
(Astrodynamical Space Test of Relativity using Optical Devices I)
mission can affect the quality of the science data as a result of
spurious Coulomb and Lorentz forces. To estimate the size of the
resultant disturbances, credible predictions of charging rates and
the charging noise are required. Using the GEANT4 software toolkit,
we present a detailed Monte Carlo simulation of the ASTROD I test
mass charging due to exposure of the spacecraft to galactic
cosmic-ray (GCR) protons and alpha particles ($^{3}$He, $^{4}$He) in
the space environment. A positive charging rate of 33.3 e$^{+}$/s at
solar minimum is obtained. This figure reduces by 50\% at solar
maximum. Based on this charging rate and factoring in the
contribution of minor cosmic-ray components, we calculate the
acceleration noise and stiffness associated with charging. We
conclude that the acceleration noise arising from Coulomb and
Lorentz effects are well below the ASTROD I acceleration noise limit
at 0.1 mHz both at solar minimum and maximum. The coherent Fourier
components due to charging are investigated, it needs to be studied
carefully in order to ensure that these do not compromise the
quality of science data in the ASTROD I mission.
\end{abstract}


\section{Introduction}

The ASTROD I (Astrodynamical Space Test of Relativity using Optical
Devices I) mission concept is a down-scaled version of ASTROD. The
main objectives of ASTROD I are: to improve the precision of
measurement of solar-system dynamics, solar-system constants and
ephemeris; to measure the relativistic gravitational effects; to
test the fundamental laws of space-time more precisely and to
improve the measurement of the rate of change of the gravitational
constant with time.\cite{Ni1}

  The basic scheme of the ASTROD I space mission is to use two-way laser
interferometric ranging and laser pulse ranging between a drag-free
ASTROD I spacecraft in a solar orbit and deep space laser stations
on Earth. The ASTROD I spacecraft is 3-axis stabilized with a total
mass of 300-350 kg. The mass of the payload is 100-120 kg. The
science data rate is 500 bps. The spacecraft is cylindrical with OD
(Outer Diameter) 2.5 m and height 2 m and has its side surface
covered with solar panels. In orbit, the cylindrical axis will be
perpendicular to the orbit plane with the telescope pointing toward
the ground laser station. The effective area to receive sunlight is
about 5 m$^{2}$ and can generate over 500 W of
power.\cite{Ni1}\cdash\cite{Ni2}

  The spacecraft will be launched into the solar orbit from a low earth orbit.
The injection correction will be made using a medium-sized ion
thruster. A launch on August 4, 2010 would provide a suitable orbit.
The orbit in the X-Y plane of the heliocentric ecliptic coordinate
system is shown in Figure 1.\cite{Ni1}$^{,}$\cite{Tang} This solar
orbit will initially have a period of 290 days. After two
gravity-assist encounters with Venus, the period will be shortened
to about 165 days. After about 370 days from launch, the spacecraft
will arrive at the other side of the Sun. The spacecraft will have
the first closest approach to Venus 107.8 days after launch with a
distance of 31606.0 km to the centre of Venus. The spacecraft
crosses the Venus trajectory in front of Venus and gets a swing
toward the Sun to achieve the Venus orbital period. After the first
encounter, the spacecraft has the same orbit's period as Venus and
encounters Venus again after about 1 period (224.7 days) with a
closest approach distance of 16151.7 km from the centre of
Venus.\cite{Ni1}\cdash\cite{Ni2}
\begin{figure}[]\vspace*{-22pt}
\centerline{\psfig{file=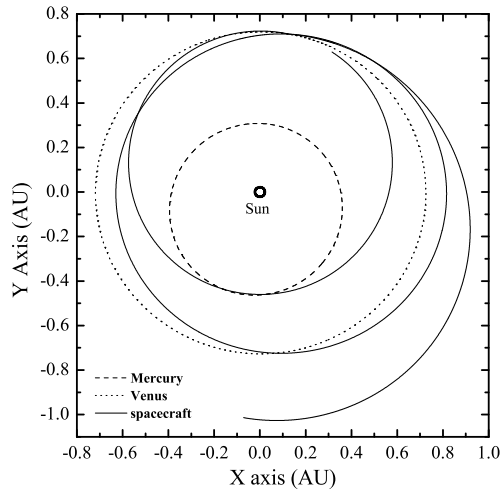,width=7.7cm}} \vspace*{-16pt}
\caption{The ASTROD I orbit in the X-Y plane of the heliocentric
ecliptic coordinate system.\label{f1}}
\end{figure}

  To achieve its goal, the ASTROD I residual acceleration noise target is:
\begin{equation}
S_{\triangle{a}}^{1/2}(f) = 3\times10^{-14}[\frac{0.3\
\mathrm{mHz}}{f}+30(\frac{f}{3\ \mathrm{mHz}})^{2}]\
\mathrm{ms^{-2}Hz^{-1/2}}, \label{eqn1}
\end{equation}
over the frequency range of 0.1 mHz $< f <$ 100 mHz.\cite{Ni2} Here
$S_{\triangle{a}}^{1/2}(f)$ is the residual acceleration noise
spectral density. It is compared to the LISA Pathfinder LISA
Technology Package,\cite{Vitale} LISA\cite{Bender} and
ASTROD\cite{Ni3} noise target curves in Figure 2.
\begin{figure}[]\vspace*{-20pt}
\centerline{\psfig{file=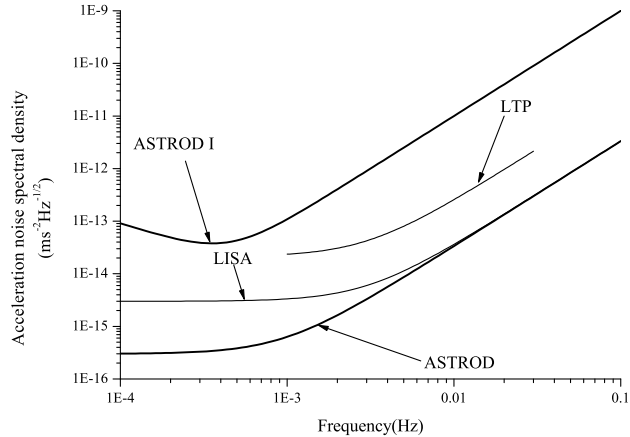,width=9.7cm}} \vspace*{-18pt}
\caption{A comparison of the target acceleration noise curves of
ASTROD I, the LTP, LISA and ASTROD. \label{f2}}
\end{figure}

  The test mass is key to guarantee the
drag-free condition for ASTROD I. It is a 1.75 kg, rectangular
parallelepiped, made of an extremely low magnetic susceptibility ($<
5\times10^{-5}$) Au-Pt alloy, to minimize magnetic disturbances. The
test mass is placed in the centre of spacecraft and surrounded by
electrodes on all six sides. Any relative displacement of test mass
to the surrounding electrodes will be capacitively detected and the
spacecraft will follow it using FEEP (Field Emission Electric
Propulsion) to ensure drag-free flight. Cosmic rays and solar
energetic particles will easily penetrate the light shielding of the
spacecraft to transfer heat, momentum and electrical charge to the
test mass. Electrical charging is the most significant of these
disturbances. It will result in forces on the test mass, due to
Coulomb and Lorentz interactions, which will disturb the geodesic
motion. The characteristics of the test mass charging process depend
on the incident flux, spacecraft geometry and the physical processes
that occur. The three main disturbances associated with this charge
are an increase in the test mass acceleration noise, coupling
between the test mass and the spacecraft and the appearance of
coherent Fourier components in the measurement
bandwidth.\cite{Shaul1} To limit the acceleration noise associated
with Coulomb and Lorentz forces to meet the ASTROD I noise
requirement, the test mass must be discharged in orbit. Our previous
work predicted the charging rates for ASTROD I test mass from
galactic cosmic rays at solar minimum using a simplified geometry,
and using these predictions, estimated the magnitude of disturbances
associated with charging.\cite{Bao1} In this paper, we present the
detailed calculation of the ASTROD I net test mass charging rate and
shot noise, due to cosmic rays at solar minimum and solar maximum,
with a more realistic geometry model. Based on these results, we
estimate the magnitude of acceleration noise, stiffness and the
coherent signals associated with charging.

\section{Modelling the Charging Process}

\subsection{Radiation environment model}

We have simulated the fluxes of the 3 most abundant primaries,
proton, $^{3}$He and $^{4}$He primary particles which represent
approximately 98\% of the total cosmic ray flux. Near-Earth cosmic
ray spectra were adopted, as used in similar LISA simulations. These
are shown in Figure 3.\cite{Grimani1}
\begin{figure}[]\vspace*{-20pt}
\centerline{\psfig{file=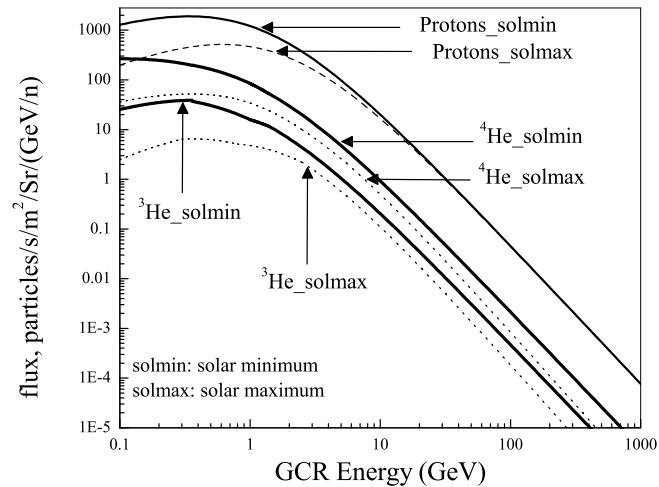,width=9.7cm}} \vspace*{-18pt}
\caption{Differential energy spectra for cosmic ray protons and He
nuclei at near earth orbit. For each species, the upper curve
indicates the solar minimum spectrum, the lower curve indicates
the solar maximum spectrum. \label{f3}}
\end{figure}
The primary particles are generated from points sampled uniformly
from a spherical surface of 3000 mm diameter, encompassing the whole
ASTROD I geometry model.

  The total time $T$ for
bombardment by $N_{0}$ primaries is given by $T = N_{0}/F\cdot\pi
R^{2}$, where $F$ is the integral flux (unit: particles/cm$^{2}$/s)
for each species at solar minimum or solar maximum, and  $R = 1500 $
mm. The $N^{th}$ particle event occurs at time $t = N\cdot
T/N_{0}$.\cite{Araujo1} The effects on charging of other particle
species (C, N, O, e$^{-}$) are determined separately, based on a
LISA study.\cite{Grimani2} Work is under way to study the impact on
the charging disturbances from variations in the incident flux due
to the ASTROD I heliocentric position changes.

\subsection{Physics model}

Test mass charging depends heavily on the physics processes that
occur during the passage of the particles through matter and the
geometry model used in simulation. The GEANT4 toolkit employs Monte
Carlo particle ray-tracing techniques to follow all primary and
secondary particles. Due to their high energy and hadronic nature,
cosmic rays can produce the complex nuclear reactions which have
large final-state multiplicities, producing many secondary
particles. A low energy threshold of 250 eV was imposed for
secondary particle production in our simulation. The physics
processes simulated include electromagnetic, hadronic and
photonuclear interactions. Fluorescence and non-radiative (Auger)
atomic deexcitation have been implemented. The hadronic physics is
mainly implemented by elastic and inelastic scattering processes.
The inelastic reactions were based on the LEP (Low Energy Particles)
and HEP (High Energy Particles) parameterized models. The inelastic
reactions also use evaporation models to treat the deexcitation of
nuclei with A $>$ 16, comprising gamma emission, fragment
evaporation (p, n, $\alpha$, $^{2}$H and $^{3}$H) and fission of
heavier residual nuclei. A variety of decay, capture and
annihilation processes has also been included in our physics
processes list based on LISA GEANT4 model of Araujo et
al.\cite{Araujo2} The charging potential of several additional
physics processes, such as the kinetic emission of very low energy
electrons which has not been modeled in the present simulation, has
been assessed based on LISA studies.\cite{Araujo1}

\subsection{Geometry model}
The basic payload configuration of ASTROD I is sketched in Ref. 1.
The geometry model built for ASTROD I using GEANT4 in present work
is sketched in Figure 4 and Figure 5. Table 1 lists the dimensions
and weight of main constituent parts of the ASTROD I geometry model.
Table 2 lists the composition and density of materials used in the
model.

The cylindrical structure of the spacecraft consists of a layer with
diameter 2.5 m, height 2 m and thickness 10 mm made of CFRP (Carbon
Fibre Reinforced Plastic) honeycomb. The top and bottom of the
spacecraft are covered by the upper deck and lower deck. To prevent
sunlight from striking the inside directly and to reduce the
temperature perturbation inside, all surfaces of spacecraft are
covered by a thermal shield consistent of five layers of materials
(face sheet, honey comb core, face sheet, foam, face sheet). The
edges of upper deck and thermal shield are shown as large ellipses
in Figure 4. The inner lower deck is shown as the grey part in
Figure 4 and Figure 5. The payload structure is used for shielding
the optical bench, inertial sensor and primary telescope etc. The
primary telescope, which collects the incoming light is a 500 mm
diameter f/1 Cassegrain telescope.\cite{Bao2} Some 30 boxes
represent the components above $\sim$ 0.1 kg based on the payload
configuration.\cite{Ni1} The mass of the spacecraft and payload are
estimated as 341 kg and 109 kg in this study.
\newpage
\begin{figure}[]\vspace*{0pt}
\centerline{\psfig{file=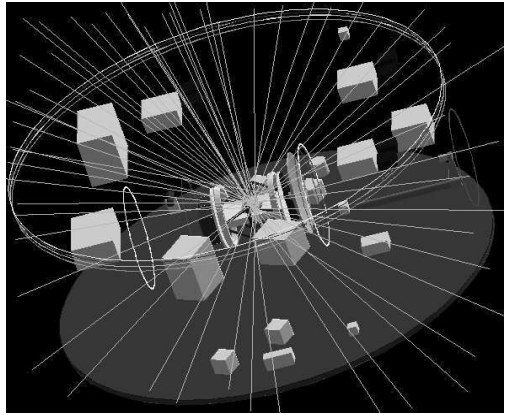,width=6.7cm}} \vspace*{0pt}
\caption{The schematic diagram for the geometry model with a
simulated GEANT4 cosmic-ray event. \label{f4}}
\end{figure}

\begin{figure}[]\vspace*{-16pt}
\centerline{\psfig{file=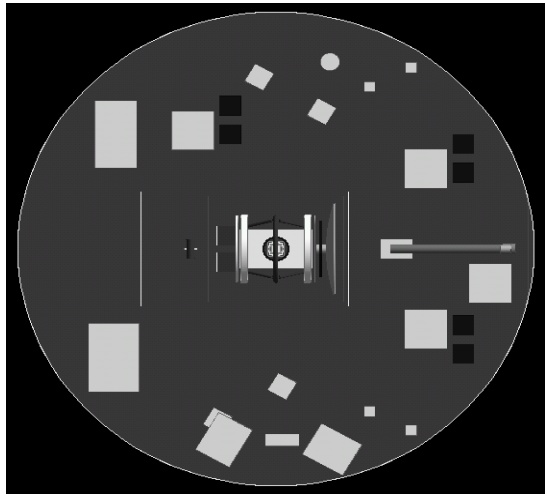,height=6.6cm,width=7.2cm}}
\vspace*{0pt} \caption{ The overhead view of the GEANT4 geometry
model for ASTROD I. The six black boxes are laser heads and the
other white boxes represent the components above $\sim$ 0.1 kg.
\label{f5}}
\end{figure}

\newpage
\begin{table}[ph]\vspace*{0pt}
\tbl{Dimension and weight of the main constituent parts of the
ASTROD I geometry model}
{\begin{tabular}{|l|l|c|l|l|}
 \hline
  Constituent part & Material & Weight    & Dimension (mm)                    & Comment\\
  &&(kg)&&\\
\hline Solar Panels  & Scell & $\sim$61.5     & Tube: radius 1250;                & covers the side \\
                     &       &                & height 2000; thickness 0.5        & surface of the\hphantom{00}\\
                     &&&&spacecraft\\
\hline Spacecraft    & CFRP  & $\sim$7.8      & Tube: radius 1250;                & contains the \\
                     &       &                & height 2000; thickness 10        & payloads\\
\hline Thermal Shield& CFRP  & $\sim$100.3   & Tube: radius 1240;                & covers the \\
                     & Al-Honeycomb  &   & height 2000;      & spacecraft\\
&foam&& thickness 41.8 &\\
\hline Upper Deck    & CFRP  & $\sim$15       & Cylinder: radius 1240;                        & top of the \\
                     & Al-Honeycomb &         & thickness 30                    & spacecraft\\
\hline Lower Deck    & CFRP  & $\sim$15       & Cylinder: radius 1240;            & above the \\
                     & Al-Honeycomb &         & thickness 30                      & MLI-blanket\\
\hline  MLI-blanket  & MLImat  & $\sim$6.8    & Cylinder: radius 1240;                          & bottom of the \\
                     &              &         & thickness 1     & spacecraft\\
\hline  Payload Shield & CFRP  & $\sim$15.6   & Tube: radius 300;                                     & Shields the \\
                       &  &                   & length 1000; thickness 5                                           & optical bench, \\
                       &&&& inertial sensor\\
\hline  Optical Bench  & ULEglass  & $\sim$5  & Rectangular: length 350; & contains the \\
                       &           &      &     width 200; height 40  & Ti-house \\
\hline  Ti House       & Ti Alloy  & $\sim$1.9  & Tube: radius 62.5;           & located in the  \\
                       &           &       &    height 224; thickness 5        & optical bench\\
\hline  Mo House       & Molybdenum  & $\sim$2.5  & Cube: length 75 & contains the \\
&&&& test mass\\
\hline  Test Mass      & AuPt Alloy  & $\sim$1.75  & Rectangular: length 50;  & \\
                       &           &       & width 50; height 35    & \\
\hline  Primary  & SiC  &  $\sim$6.4   & Dish: radius 250;&  collects the \\
          Telescope             &           &       &      thickness 10.5                    & incoming light \\
\hline  Shield/Mounting  & CFRP  & $\sim$1.6  & Cylinder: radius 175; & \\
         Plate       &              &         & thickness 10     & \\
\hline  Telescope Shield & CFRP  & $\sim$12.6     & Tube: radius 285; & Shields the  \\
                       &  &              &  length 860; thickness 5   &  secondary mirror \\
\hline  PCDU  & Al6061  & $\sim$15.9  & Rectangular: length 350; & Power Condition-\\
            &           &      &  width 200; height 300    &  ing and \\
            &&&&Distribution Unit\\
\hline  Transponder  & Al6061  & $\sim$3.5  & Rectangular: length 220; & \\
             &           &                      & width 184; height 178    & \\
\hline  CPS   & Al6061  & $\sim$15.9  & Rectangular: length 240; & Centralised \\
                       &           &      & width 356; height 140      & Processor System\\
\hline  Interferometer   & Al6061  & $\sim$3.5  & Rectangular: length 200;& \\
    Electronic Boxes                   &           &     &    width 200; height 150  & \\
\hline  Gyroscope   & Al6061  & $\sim$1  & Cylinder: radius 42.5; & \\
                       &           &      &  height 89    & \\
\hline  RFDU   & Al6061  & $\sim$1  & Rectangular: length 160; & Radio Frequency\\
            &           &      & width 60; height 80     & Distribution Unit\\
\hline
\end{tabular}\label{ta1}
}
\end{table}
\newcommand{\newhao}{\fontsize{7.7pt}{\baselineskip}\selectfont}

\newpage
\begin{table}[ph]
\tbl{The composition and density of materials used in the ASTROD I
geometry model (CFRP:Carbon Fibre Reinforced Plastic)}
{\begin{tabular}{|c|c|c|c} \hline
  Material & Composition (by weight) & Density (g/cm$^{3}$) \\
\hline   Vacuum     & gas & 1.0$\times$10$^{-25}$\\
\hline   Al6061     & Al(98\%), Mg(1\%), Si(0.6\%), Fe(0.4\%)  & 2.70\\
\hline   Al Honeycomb     & Al(98\%), Mg(1\%), Si(0.6\%), Fe(0.4\%)  & 0.05\\
\hline   MLImat     & H(4.1958\%), C(62.5017\%), O(33.3025\%) & 1.40\\
\hline  Ti Alloy    & Ti(90\%), Al(6\%), V(4\%) & 4.43 \\
\hline  AuPt Alloy  & Au(70\%), Pt(30\%)  & 19.92\\
\hline  ULE Glass   & SiGlass(92.5\%), TiGlass(7.5\%) & 2.21\\
\hline  foam        & C(90\%), H(10\%) & 0.05\\
\hline  SiGlass     & O: 2, Si: 1 & 2.20\\
\hline  TiGlass     & O: 2, Ti: 1 & 4.25\\
\hline  SHAPAL      & Al: 2, N: 1 & 2.90\\
\hline  SiC         & Si: 1, C: 1 & 3.10\\
\hline   Scell      & Si & 7.82\\
\hline  Molybdenum  & Mo & 10.22\\
\hline  Gold        & Au & 19.32\\
\hline  CFRP        & C & 1.66\\
\hline  carbon      & C & 2.10  \\
\hline  CFRP Honeycomb & C & 0.05 \\
\hline
\end{tabular}
}
\end{table}

  A $50\times 50\times 35$ mm$^{3}$ test mass is
at the centre of the spacecraft. The test mass is housed inside
capacitance sensors located in optical bench mounted behind the
telescope. The test mass is surrounded by sensing and actuation
electrodes lodged in a molybdenum housing. A 0.3\ $\muup$m gold
layer is plated on the entire inner surface of the sensor housing.
The assembly is accommodated in a titanium vacuum ($< 10\ \muup$Pa)
enclosure. The gap between test mass and electrodes along X axis or
Y axis is 4 mm; that along Z axis is 2 mm.\cite{Bao2} The GEANT4
model for inertial sensor is shown in Figure 6. \vspace*{-5pt}
\begin{figure}[]
\centerline{\psfig{file=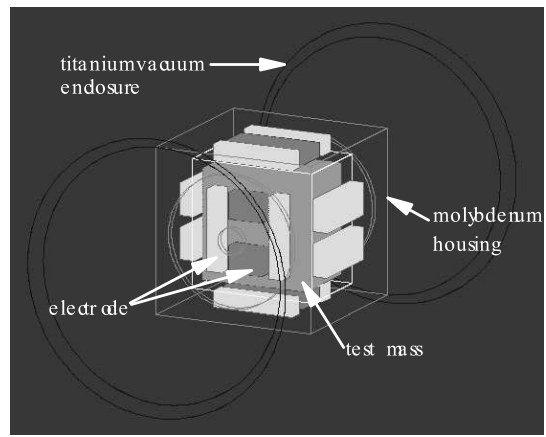,width=7.2cm}} \vspace*{-1pt}
\caption{ASTROD I inertial sensor model implemented in Geant4. The
test mass, located at the centre of the figure, is surrounded by
sensing electrodes (white) and injection electrodes (grey).
\label{f6}}
\end{figure}
\newpage
\section{Charging Results}
We have run 6 independent GEANT4 simulations to determine the
charging of the ASTROD I test mass by cosmic ray protons, $^{3}$He
and $^{4}$He, at solar minimum and maximum. In total, about
8,500,000 events were simulated. The details of each event that
resulted in test mass charging were recorded, including the event
time, net charge deposited on the test mass and the energy of the
primary.
\subsection{Charging simulation results}
The variation of the net test mass charge with time, due to GCR
proton, $^{3}$He and $^{4}$He fluxes are shown respectively in
Figure 7, Figure 8 and Figure 9 at solar minimum and maximum. The
straight lines in these figures correspond to least squares fits of
the simulated data, giving mean net charging rates attributable to
the proton, $^{3}$He and $^{4}$He fluxes. The proton flux is
responsible for positive charging rates of 26.5 $\pm$ 0.5 e$^{+}$/s
at solar minimum and 9.0 $\pm$ 0.5 e$^{+}$/s at solar maximum; The
$^{3}$He flux is responsible for positive charging rates of 0.8
$\pm$ 0.05 e$^{+}$/s at solar minimum and 0.3 $\pm$ 0.05 e$^{+}$/s
at solar maximum; The $^{4}$He flux is responsible for positive
charging rates of 6.0 $\pm$ 0.20 e$^{+}$/s at solar minimum and 2.4
$\pm$ 0.20 e$^{+}$/s at solar maximum. The uncertainties quoted are
only associated with the Monte Carlo fluctuations. Our simulation
indicates that $\sim$ 97\% of the charge accumulated comes from
primary cosmic ray protons and $^{4}$He at both solar minimum and
maximum and all three fluxes lead to positive charging of the test
mass. The proton flux dominates these rates. However, $^{4}$He,
which constitutes only 8\% of the total cosmic rays flux, is
responsible for $\sim$ 18\% of the test mass charging at solar
minimum and $\sim$ 20\% at solar maximum.

\vspace*{-12pt}
\begin{figure}[]
\centerline{\psfig{file=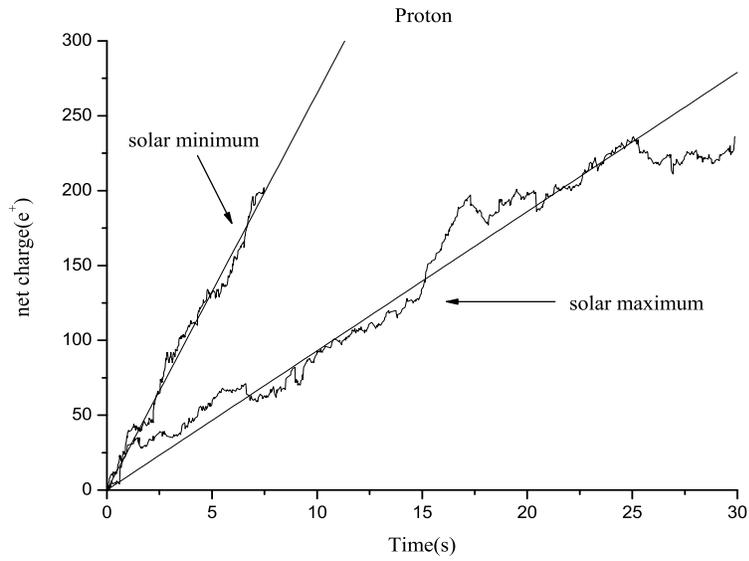,height=8.6cm,width=11.2cm}}
\vspace*{-18pt} \caption{The charging timeline for protons at solar
minimum and maximum. The straight line is a least squares fit.
\label{f7}}
\end{figure}

\newpage
\begin{figure}[]\vspace*{-10pt}
\centerline{\psfig{file=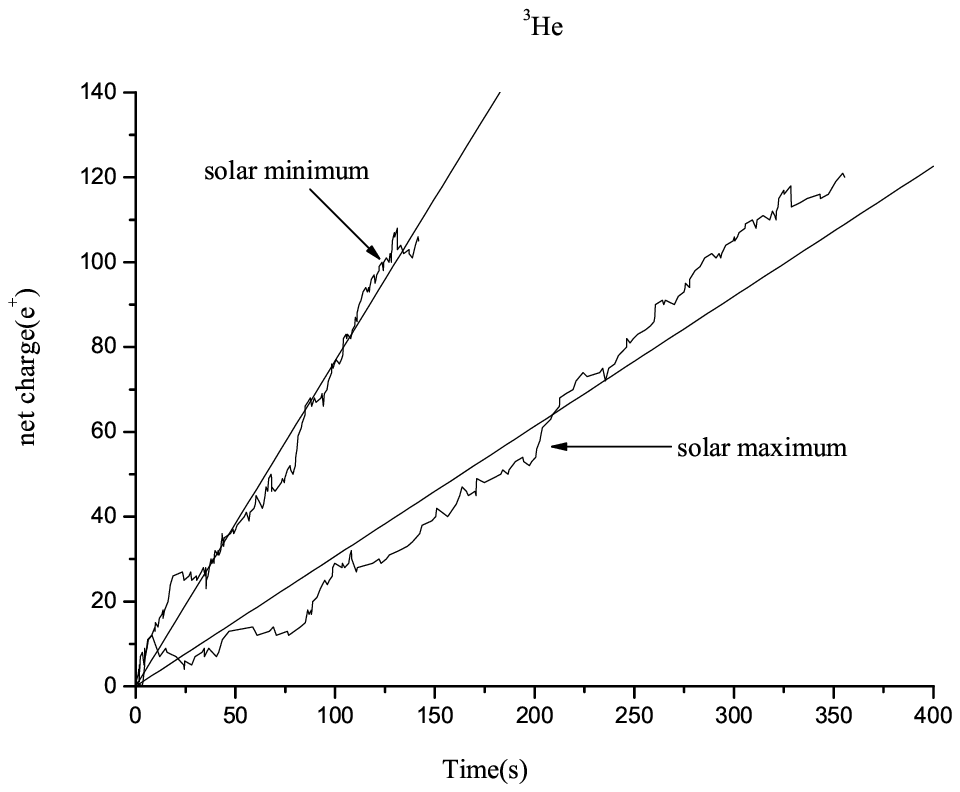,height=8.6cm,width=11.2cm}}
\vspace*{-18pt} \caption{The charging timeline for $^{3}$He at solar
minimum and maximum. The straight line is a least squares fit.
\label{f8}}
\end{figure}

\newpage

\begin{figure}[]\vspace*{-28pt}
\centerline{\psfig{file=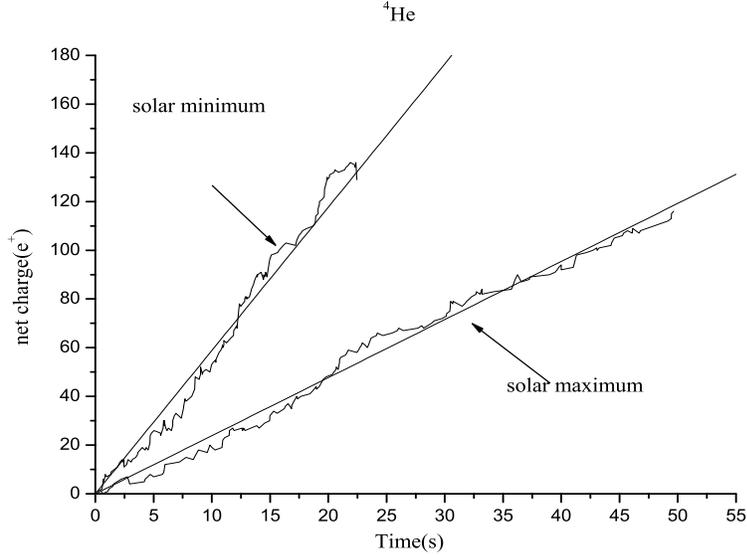,height=8.6cm,width=11.2cm}}
\vspace*{-18pt} \caption{The charging timeline for $^{4}$He at solar
minimum and maximum. The straight line is a least squares fit.
\label{f9}}
\end{figure}

  Two histograms of the net charge deposited in an event are given
in Figure 10 and Figure 11, for the proton data set, showing that
most events result in the transfer of one unit of charge. The
effects of the positive and negative chargings cancel to some
extent. An imbalance in these currents gives rise to the net
positive charging rate.

\begin{figure}[]\vspace*{-42pt}
\centerline{\psfig{file=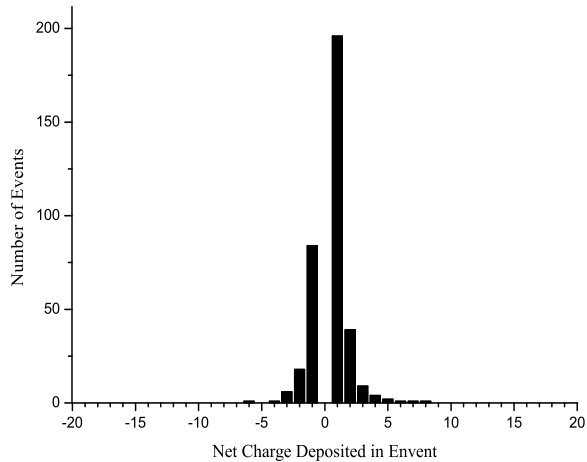,height=9.4cm,width=10.8cm}}
\vspace*{-40pt} \caption{Histogram of the net charge deposited in an
event, for incident protons at solar minimum. The total number of
proton events simulated was 2,290,000.  \label{f10}}
\end{figure}

\begin{figure}[]\vspace*{-52pt}
\centerline{\psfig{file=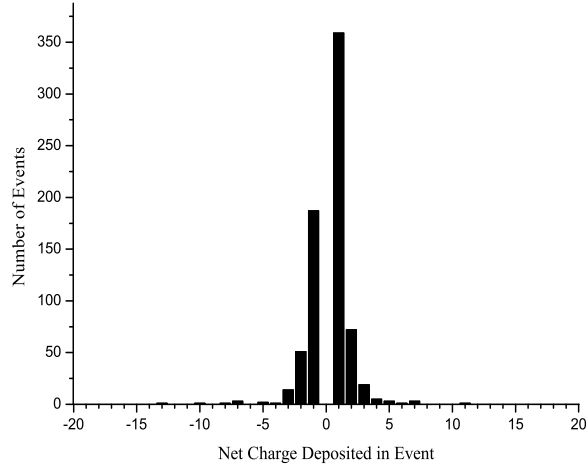,height=9.4cm,width=10.8cm}}
\vspace*{-42pt} \caption{Histogram of the net charge deposited in an
event, for incident protons at solar maximum. The total number of
proton events simulated was 4,000,000.\label{f11}}
\end{figure}
\vspace*{-8pt}
  The charging rate is plotted as a function of primary energy in
Figure 12. The low energy cut-off is due to the shielding provided
by the spacecraft, which prevents incident protons with energies
below $\sim$ 100 MeV from charging the test mass. At solar minimum,
the most significant charging mechanism is primary cosmic ray
particles stopping in the test mass. This occurs mainly for protons
of energy between $\sim$ 100 - 720 MeV. Protons with energies in
excess of $\sim$ 720 MeV have sufficient energy to traverse the
distance through the spacecraft to the test mass and the longest
path through the test mass, without being stopped. This explains the
peak observed in this energy interval in Figure 12 at solar minimum.
The scenario at solar maximum is distinct: a peak is visible at
higher energies because the primary proton flux peak shifts towards
higher energy.\vspace*{-10pt}

\begin{figure}[]\vspace*{-26pt}
\centerline{\psfig{file=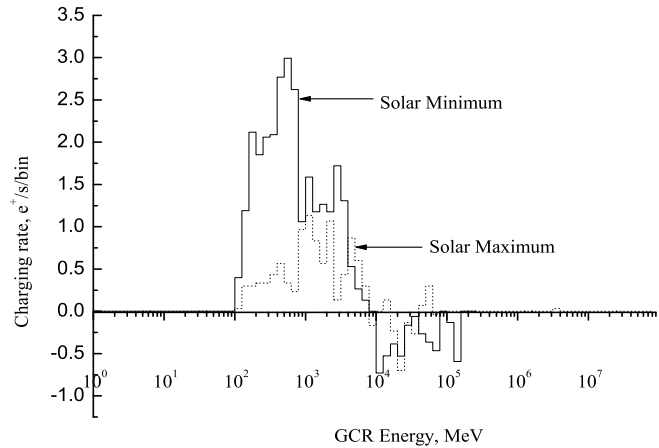,height=7.0cm,width=9.8cm}}
\vspace*{-18pt} \caption{Charging rate of test mass as a function of
primary proton energy at solar minimum and at solar
maximum.\label{f12}}
\end{figure}

\newpage
  In addition to the Monte Carlo uncertainty, we added an error of
$\pm$30\% in the net charging rates to account for uncertainties in
the GCR spectra, physics models and geometry implementation.
Further, based on LISA studies\cite{Araujo1}$^{,}$\cite{Grimani2}, a
potential contribution to the charging rate of 28.4 e$^{+}$/s at
solar minimum and of 17.0 e$^{+}$/s at solar maximum from kinetic
low energy secondary electron emission should be considered; the
effect of cosmic ray fluxes of particle species not included in this
simulation is expected to increase the net charging rate by $\sim$
4.2\% at solar minimum and $\sim$ 7.3\% at solar maximum\cite{Bao2}.

\subsection{Charging noise}
Following to Ref. 12, the charging flux is considered to be made up
of independent currents, $I_{q}$, each composed solely of charges
$qe$ ($q$=+1 for protons), with shot noise of single-sided spectral
density $S_{q}=\sqrt{2qeI_{q}}$, where $e$ is the magnitude of
electron charge. The total noise, $S_{R}$, is then given by the
quadrature sum of $S_{q}$, over all values of $q$. Considering the
Monte Carlo currents alone, $S_{q}$ = 17.6 es$^{-1}$Hz$^{-1/2}$ at
solar minimum and $S_{q}$ = 9.6 es$^{-1}$Hz$^{-1/2}$ at solar
maximum. Based on the LISA study\cite{Araujo1}, low energy secondary
electron emission is estimated to contribute an extra 10.3
es$^{-1}$Hz$^{-1/2}$ at solar minimum and 8.0 es$^{-1}$Hz$^{-1/2}$
at solar maximum; the charging noises from other species (C, N, O,
e$^{-}$) are estimated to be 8.9 es$^{-1}$Hz$^{-1/2}$ at solar
minimum and 3.7 es$^{-1}$Hz$^{-1/2}$ at solar maximum. Integrating
in the time domain gives the charging fluctuations at frequency $f$,
\begin{equation}
S_{Q}(f) = S_{R}/2{\pi}f. \label{eqn2}
\end{equation}
The charging rate and noise contributions from the different sources
mentioned above are summarized in Table 3. Given the results in
Table 3, by adding the contributions of all the independent sources,
we estimate the total worst case charging rate to be 73.8 e$^{+}$/s
at solar minimum and 33.8 e$^{+}$/s at solar maximum. The total
worst case noise is 22.3 es$^{-1}$Hz$^{-1/2}$ at solar minimum and
13.0 es$^{-1}$Hz$^{-1/2}$ at solar maximum.

\begin{table}[ph]
\tbl{The charging rate and noise contributions from different
sources.} {\begin{tabular}{@{}ccccc@{}} \toprule Source &
~~~~~~charging
rate(e$^{+}$/s) & ~~~~~~charging noise(e/s/Hz$^{1/2}$) & \\
& min ~~~~~ max & min ~~~~~~ max \\ \colrule
p\hphantom{00} & 26.5\hphantom{00}~~~~~9.0\hphantom{0} & 15.9\hphantom{00}~~~~~8.6\hphantom{0}  \\
$^{3}$He\hphantom{00} & 0.8\hphantom{0}~~~~~~0.3 & 2.5\hphantom{0}~~~~~~1.6  \\
$^{4}$He\hphantom{00} & 6.0\hphantom{0}~~~~~~2.4 & 7.2\hphantom{0}~~~~~~3.9  \\
Secondary Electron\hphantom{00} & 28.4\hphantom{0}~~~~~17.0\hphantom{0} & 10.3\hphantom{00}~~~~8.0\hphantom{0} \\
Other Species(C, N, O, e$^{-}$)\hphantom{00} & 1.4\hphantom{0}~~~~~~0.9 & 8.9\hphantom{0}~~~~~~3.7\\
Uncertainty\hphantom{00} & 10.0\hphantom{00}~~~~~3.5\hphantom{0} & -~~~~~~~~~~-\\
\botrule
\end{tabular} \label{ta1}}
\end{table}

\section{Acceleration Disturbances}

\subsection{Coulomb noise and stiffness}
The charge-dependent Coulomb acceleration $a_{ _{QK}}$ in
direction $\hat{k}$ is given by: {
\begin{equation}
 a_{_{QK}}=\frac{Q^{2}}{2mC_{T}^{2}}\frac{\partial{C_{T}}}{\partial{k}}+\frac{QV_{T}}{mC_{T}}\frac{\partial{C_{T}}}{\partial{k}}-
\frac{Q}{mC_{T}}\sum\limits _{i=1}
^{N-1}V_{i}\frac{\partial{C_{i,N}}}{\partial{k}} .\label{eqn3}
\end{equation}
} The first two terms in equation (3) are dependent on the overall
sensor geometric symmetry, through $\partial{C_{T}}/\partial{k}$,
and the third term is dependent on the symmetry of the sensor
voltage distribution.The corresponding acceleration noise, $\delta
a_{Qk}$, due to random fluctuations of the test mass position
relative to the spacecraft, $\delta k$, of the potentials of the
conductors that surround the test mass, $\delta V_{i}$, and of the
test mass free charge $\delta Q$, is given by:

\begin{equation}
 \delta{a_{_{QK}}^{2}}=(\frac{\partial{a_{ _{QK}}}}{\partial{k}})^{2}\delta{k}^{2}+\sum\limits_{i=1}^{N-1}(\frac{\partial{a_{
_{QK}}}}{\partial{V_{i}}})^{2}\delta{V_{i}}^{2}+(\frac{\partial{a_{_{QK}}}}{\partial{Q}})^{2}\delta{Q}^{2}\nonumber\\
 \label{eqn4}
\end{equation}
where $k$ is a displacement in direction $\hat{k}$; $m$ is the mass
of the test mass; $Q$ is the free charge accumulated on the test
mass; $C_{i,j}$ is the capacitance between conductors $i$ and $j$
which surround the test mass; $V_{i}$ is the potential to which
conductor $i$ is raised;
$C_{T}\equiv\sum\nolimits_{i=1}^{N-1}C_{i,N}$ is the coefficient of
capacitance of the test mass, which is defined as the $N^{th}$
conductor, with potential $V_{N}=Q/C_{T}+V_{T}$, and
$V_{T}\equiv\frac{1}{C_{T}}\sum\nolimits_{i=1}^{N-1}C_{i,N}\cdot
V_{i}$.\cite{Shaul1} The estimates for acceleration noise have
assumed typical parameter values for the ASTROD I mission: $Q$ was
taken as the amount of charge accumulated in 1 day, assuming a net
test mass charging rate of 73.8 e$^{+}$/s at solar minimum and 33.8
e$^{+}$/s at solar maximum, which corresponds to the Monte Carlo
rate, with error margins, estimated contributions from particle
species not included in the Monte Carlo model and the potential
contribution from kinetic low energy secondary electron emission,
that is likely to almost cancel in the actual sensor,\cite{Araujo1}
added; $m$ =1.75 kg; mean voltages on opposing conductors $V_{i}$ =
0.5 V; the potential difference between conductors on opposing faces
of the sensor compensated to 10 mV; the asymmetry in gap across
opposite sides of test mass is 10 $\mu$m; capacitances and
capacitance gradients were calculated using parallel plate
approximations: $C_{T}$ = 53 pF; $V_{T}$ = 0.5 V; position noise
$\delta k$ = 1$\times$10$^{-7}$ mHz$^{-1/2}$; voltage noise $\delta
V_{i}$ = 1$\times$10$^{-4}$ VHz$^{-1/2}$ and charge noise $\delta Q$
= 4.6$\times$10$^{-15}$ CHz$^{-1/2}$, which includes, as for the
charging rate, the unmodelled contributions. These Coulomb
acceleration noise due to the test mass charging are listed in Table
4. The total noise estimated here is a factor of $\sim$ 30 less than
the ASTROD I acceleration noise target. The stiffness associated
with test mass charging, $S_{Qk}$, is given by
$S_{Qk}=-m\cdot\partial a_{Qk}/\partial k$. These acceleration noise
figures are lower than the results liberally estimated by Shiomi and
Ni\cite{Shiomi} because of the smaller charging rate and charging
noise we obtained here. The requirements on Coulomb noise, stiffness
and other associated noises for ASTROD I in Ref. 14 are satisfied.
\newpage
\begin{table}[ph]
\tbl{The magnitude of charging disturbances for ASTROD I at solar
minimum and maximum, at frequency = 0.1 mHz.}
{\begin{tabular}{@{}ccccc@{}} \toprule Solar activity & Coulomb
noise & Lorentz noise & Stiffness &
\\
&($\times$10$^{-15}$ms$^{-2}$Hz$^{-0.5}$)&($\times$10$^{-15}$ms$^{-2}$Hz$^{-0.5}$)&($\times$10$^{-8}$s$^{-2}$)
\\
& $\delta k$~~~~~~$\delta V_{i}$~~~~~~$\delta Q$~~~~~~total \\
\colrule
minimum\hphantom{00} &1.06~~~~~2.18~~~~~1.32~~~~~2.80 & 2.80 & -1.52 \\
\colrule
maximum\hphantom{00} &0.49~~~~~1.08~~~~~0.77~~~~~1.40 & 1.30 & -0.69\\
\botrule
\end{tabular} \label{ta1}}
\end{table}
\vspace*{-18pt} $\delta k$: displacement noise; $\delta V_{i}$:
voltage noise; $\delta Q$: charge noise.

\subsection{Lorentz noise}
Lorentz effects arise from the motion of the test mass through the
interplanetary magnetic field, $\vec{B_{_{I}}}$ and its residual
motion through the field generated within the spacecraft,
$\vec{B_{_{S}}}$. The test mass will be housed in a conducting
enclosure, which will reduce the effect of the interplanetary
field, via the Hall effect, with efficiency $\eta$. Hence, to
first order, the Lorentz acceleration noise, $a_{_{L}}$, is given
by:

\begin{equation}
m^{2}(a_{_{L}})^{2}=(\eta{Q}V_{_{I}}\delta{B_{_{I}}})^{2}+
(\eta{Q}\delta{V_{_{I}}}B_{_{I}})^{2}+
(Q\delta{V_{_{S}}}B_{_{S}})^{2}+(\eta\delta{Q}V_{_{I}}B_{_{I}})^{2}.\label{eqn5}
\end{equation}
where $V_{I}$ is the speed of the test mass through the
interplanetary field; $\delta V_{I}$ and $\delta V_{S}$ are the
magnitudes of random fluctuations in the test mass velocity through
the interplanetary field and relative to the spacecraft,
respectively and $\delta B_{I}$ gives the magnitude of fluctuations
in the interplanetary field.\cite{Shaul1} $a_{_{L}}$ also increases
with decreasing frequency, and is estimated to be $\sim$
2.8$\times$10$^{-15}$ ms$^{-2}$Hz$^{-1/2}$ (0.1 mHz) at solar
minimum and $\sim$ 1.3$\times$$^{-15}$ ms$^{-2}$Hz$^{-1/2}$ (0.1
mHz) at solar maximum, which is a factor of $\sim$ 30 below the
ASTROD I acceleration noise target. We have assumed that $Q$ = 73.8
e$^{+}$/s at solar minimum and 33.8 e$^{+}$/s at solar maximum, as
in section 4.1; $\eta$ = 0.1; $\vec{V_{_{I}}}$ = 4$\times$10$^{4}$
m/s; $\delta V_{_{I}}$= 4.78$\times$10$^{-12}$ ms$^{-1}$Hz$^{-1/2}$;
$\delta V_{_{S}}$ = 6.28$\times$10$^{-11}$ ms$^{-1}$Hz$^{-1/2}$;
$\vec{B_{_{S}}}$ = 9.6$\times$10$^{-6}$ T; $\mid\delta B_{_{S}}\mid$
= 1$\times$10$^{-7}$ THz$^{-1/2}$; $\vec{B_{_{I}}}$=
1.2$\times$10$^{-7}$ T (this is a conservative estimate of the field
at 0.5 AU, used to give the worst-case noise, for the ASTROD I
orbit) and $\mid\delta B_{_{I}}\mid$ = 1.2$\times$10$^{-6}$
THz$^{-1/2}$.  The estimates of acceleration noises and stiffness
due to the Coulomb and Lorentz effects are listed in Table 4.

\subsection{Coherent Fourier components}

The charging of the test mass may also result in the appearance of
unwanted, coherent Fourier components in the ASTROD I measurement
bandwidth through Coulomb and Lorentz interactions, due to the time
dependence of the amount of charge accumulated on the test mass. It
is shown that the signals associated with Lorentz interactions are
expected to fall below the ASTROD I test mass residual acceleration
noise, but the signals from Coulomb interactions may exceed both the
instrumental noise over a fraction of the bandwidth, and may not be
eliminated by daily discharging of the test mass. The free charge on
the test mass at time $t$ can be expressed as follows:

\begin{equation}
Q(t)\approx \overline{\dot{Q}}t.\label{eqn6}
\end{equation}
where $t$ is the time for which the test mass has been allowed to
charge and $\overline{\dot{Q}}$ is the mean charging rate. Here,
$\overline{\dot{Q}}$ is assumed to be constant. Substituting
$Q(t)\approx \overline{\dot{Q}}t$ into the expressions for the
Coulomb and the Lorentz accelerations gives the terms\cite{Shaul2}:

\begin{equation}
f_{k}(t)\equiv\Xi_{k}t^{2}\equiv
\frac{\overline{\dot{Q}}^{~2}}{2mC_{_{T}}^{2}}\frac{\partial
C_{_{T}}}{\partial k}t^{2},\label{eqn7}
\end{equation}

\begin{equation}
e_{k}(t)\equiv \Theta_{k}t\equiv -\frac{\partial V_{_{T}}}{\partial
k}\frac{\overline{\dot{Q}}}{m}t,~~~~~~\label{eqn8}
\end{equation}

\begin{equation}
l_{k}(t)\equiv \Phi_{k}t\equiv \frac{\eta
\overline{\dot{Q}}t}{m}(\vec{V_{_{I}}}\times \vec{B_{_{I}}})\cdot
\hat{k},\label{eqn9}
\end{equation}
where $f_{k}(t)$ is dependent on the overall sensor geometric
symmetry; $e_{k}(t)$ is dependent on the symmetry of the sensor
voltage distribution; $l_{k}(t)$ is caused by the interplanetary
magnetic field. The Fourier transforms of these signals with
fixed-interval discharge will be described by a series of sinc
functions ($sinc(x)=sin\pi x/\pi x$). The equivalent one-sided power
spectral density for these coherent signals is given by
$P_{FT}^{2}=2{FT}^{2}/\tau$, where $FT$ = Fourier transform of the
signal and $\tau$ is the length of the data sample time.

Implementing the parameter values given in section 4.1 and 4.2,
taking the mean charging rate as constant and assuming that the test
mass is discharged once every 24 hours (as described in Ref. 15),
the spectral densities of $f_{k}(t)$, $e_{k}(t)$ and $l_{k}(t)$ are
estimated. The curves in figure 13 and figure 14 trace
$P_{F_{k}(f)}$, $P_{E_{k}(f)}$ and $P_{L_{k}(f)}$, at the primary
peaks of the sinc functions, where $F_{k}(f)$, $E_{k}(f)$ and
$L_{k}(f)$ are the Fourier transform of $f_{k}(t)$, $e_{k}(t)$ and
$l_{k}(t)$. $f_{k}(t)$ and $e_{k}(t)$ exceed the ASTROD I
acceleration noise limit of $3\times10^{-14}[0.3\
\mathrm{mHz}/f+30(f/3\ \mathrm{mHz})^{2}]\
\mathrm{ms^{-2}Hz^{-1/2}}$ in a fraction of frequency bins in the
frequency range of 0.1 mHz $< f <$ 4 mHz at solar minimum (see
Figure 13); only $e_{k}(t)$ exceeds the limit in a fraction of
frequency bins in the frequency range of 0.1 mHz $< f <$ 2 mHz at
solar maximum (see Figure 14). These effects are more severe as
frequency decreases. Several schemes could be used to minimize a
potential loss of the ASTROD I science data, including continuously
discharging the test mass, minimizing sensor voltage and geometrical
offsets and through spectral analysis.\cite{Shaul2}

Variations in, for example, the mean charging rate, could result in
these signals exceeding the noise target in a larger fraction of the
bandwidth. Hence, variations in these signals need to be studied
carefully as they will influence the accuracy with which the
solar-system and relativistic parameters can be determined.

A charge management system has been developed by Imperial College
London, for LISA Pathfinder. This system has been extensively
tested, both via simulations and laboratory
tests.\cite{Dis1}\cdash\cite{Shaul3} A similar system could easily
be used for ASTROD I.

\newpage
\begin{figure}[]\vspace*{-18pt}
\centerline{\psfig{file=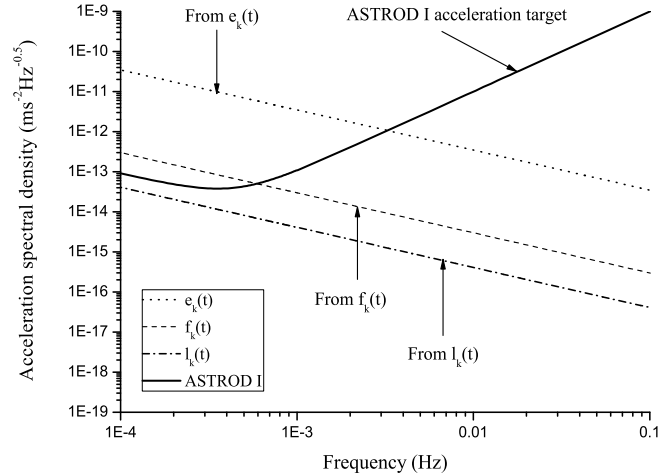,height=7.4cm,width=9.8cm}}
\vspace*{-16pt} \caption{The curves trace the spectral densities at
the primary peaks of the sinc functions at solar minimum, of the
coherent Fourier components, for $\tau$=1 year: $e_{k}(t)$ is given
by the dotted line, $f_{k}(t)$ is given by the dashed line and
$l_{k}(t)$ is given by the dashed-dotted line. The bold full line
gives the ASTROD I acceleration noise limit. \label{f13}}
\end{figure}

\begin{figure}[]\vspace*{-18pt}
\centerline{\psfig{file=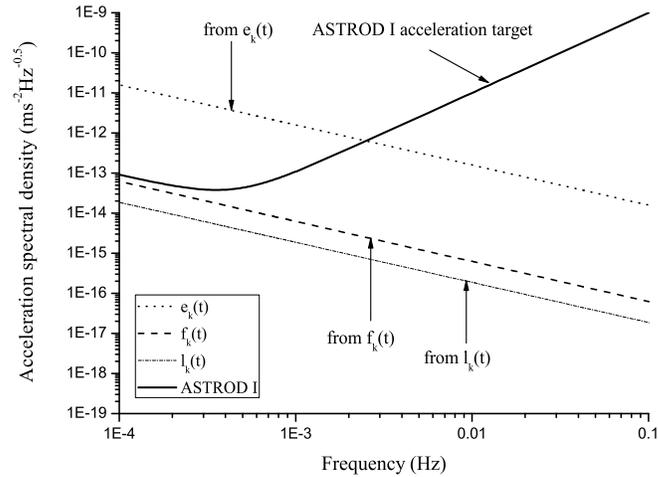,height=7.4cm,width=9.8cm}}
\vspace*{-16pt} \caption{The curves trace the spectral densities at
the primary peaks of the sinc functions at solar maximum, of the
coherent Fourier components, for $\tau$=1 year: $e_{k}(t)$ is given
by the dotted line, $f_{k}(t)$ is given by the dashed line and
$l_{k}(t)$ is given by the dashed-dotted line. The bold full line
gives the ASTROD I acceleration noise limit.\label{f14}}
\end{figure}

\section{Conclusion}
The charging of the ASTROD I test mass by cosmic ray protons and
alpha particles ($^{3}$He and $^{4}$He) has been simulated using the
GEANT4 toolkit at solar minimum and maximum. The Monte Carlo model
predicted a net charging rate of $\sim$ 11.7 e$^{+}$/s at solar
maximum, rising to 33.3 e$^{+}$/s at solar minimum. Although the
proton flux is the dominant charging flux, $^{4}$He, which
constitutes only 8\% of the total cosmic ray flux, is responsible
for $\sim$ 18\% and $\sim$ 20\% of this rate at solar minimum and
maximum, respectively. We have also included an additional net
charging rate contribution due to particle species that were not
included in the Monte Carlo model, and a potential charging
mechanism, due to kinetic low energy secondary electron emission
based on LISA studies.\cite{Araujo1}$^{,}$\cite{Grimani2} There is
an additional uncertainty of $\pm$ 30\% in the net charging rate,
due to uncertainties in the cosmic ray spectra, physics models and
geometry implementation. A recent, preliminary, comparison of a
simplified GEANT4 simulation and actual GP-B charging rates,
indicate ~45\% agreement (GP-B simulation is larger than
experimental value by 45\%), reinforcing confidence in these
predictions.\cite{Shaul3}$^{,}$\cite{Turn}

  The ASTROD I acceleration noise limit is 10$^{-13}$
  ms$^{-2}$Hz$^{-1/2}$
at 0.1 mHz, which is less stringent than the LISA requirement. The
magnitudes of the Coulomb and Lorentz acceleration noise associated
with test mass charging increase with decreasing frequency. At the
lowest frequency in the ASTROD I bandwidth, 0.1 mHz, the estimates
of the Coulomb and Lorentz acceleration noise are both well below
the acceleration noise target. These results agree, to within 30\%,
with those from our earlier study\cite{Bao1}, which was based on a
simple geometry model. The variations in the test mass charging rate
will alter the spectral description of the coherent Fourier
components. Hence, further work is needed to ensure that these do
not compromise the quality of the science data of the ASTROD I
mission.

The charging process of the ASTROD I test mass by SEPs (Solar
Energetic Particles) has also been simulated using the GEANT4
toolkit, and the charging rate is much larger than the values due to
GCR (Glactic Cosmic Ray) proton flux at solar maximum and solar
minimum.\cite{Liu1}\cdash\cite{Shaul4} However, the charge
management hardware described in reference 18 could be used to
discharge a test mass even during solar events, provided safe
operation could be ensured.

The effect of cosmic ray fluxes of particle species not included in
the Monte Carlo simulation needs to be verified for the ASTROD I
geometry. According to the recent work by Grimani et
al.,\cite{Grimani3} the charging rate in each LISA test mass induced
by primary and interplanetary electrons is comparable (absolute
value) to that released by the nuclei of the C, N, O group at solar
minimum. The acceleration disturbances due to charging of the ASTROD
I test mass by electrons are also under study. Further, we will
evaluate the variation in the ASTROD I test mass charging rate over
the orbit over the solar cycle, including a detailed study of SEP
events, and its variation due to modulation of cosmic ray flux over
the ASTROD I orbit. For this, a SCoRE (Solar And Cosmic Ray Physics
And The Space Environment: Studies For And With LISA) study along
the line of Shaul et al.\cite{Shaul4} would be useful.

\section*{Acknowledgments}

This work is funded by the National Natural Science Foundation
(Grant Nos 10475114 and 10573037) and the Foundation of Minor
Planets of Purple Mountain Observatory.


\end{document}